\documentclass[twocolumn,tighten]{aastex63}
\usepackage{amssymb, amsmath,framed}

\usepackage{bm}
\expandafter\ifx\csname package@font\endcsname\relax\else
 \expandafter\expandafter
 \expandafter\usepackage
 \expandafter\expandafter
 \expandafter{\csname package@font\endcsname}
\fi
\def\bq{\begin{equation}}
\def\eq{\end{equation}}
\def\bqy{\begin{eqnarray}}
\def\eqy{\end{eqnarray}}

\def\p{\partial}

\def\p{\partial}

\def\R{\mathbb{R}}

 \def\q#1#2{q^{\hspace{1pt}#1}_{\,,\hspace{1pt}#2}}
 \def\a#1#2{a^{\hspace{1pt}#1}_{\,,\hspace{1pt}#2}}
 \def\d#1#2{\delta^{\hspace{1pt}#1}_{\,,\hspace{1pt}#2}}

\begin{document}
\title{\large{Excitation Properties of Photopigments and Their Possible Dependence on the Host Star}}

\correspondingauthor{Manasvi Lingam}
\email{mlingam@fit.edu}

\author{Manasvi Lingam}
\affiliation{Department of Aerospace, Physics and Space Sciences, Florida Institute of Technology, Melbourne, FL 32901, USA}
\affiliation{Department of Physics and Institute for Fusion Studies, The University of Texas at Austin, Austin, TX 78712, USA}

\author{Amedeo Balbi}
\affiliation{Dipartimento di Fisica, Universit\`a di Roma ``Tor Vergata", 00133 Roma, Italy}

\author{Swadesh M. Mahajan}
\affiliation{Department of Physics and Institute for Fusion Studies, The University of Texas at Austin, Austin, TX 78712, USA}
\affiliation{Department of Physics, School of Natural Sciences, Shiv Nadar University, Greater Noida, Uttar Pradesh, 201314, India}

\begin{abstract}
Photosynthesis is a plausible pathway for the sustenance of a substantial biosphere on an exoplanet. In fact, it is also anticipated to create distinctive biosignatures detectable by next-generation telescopes. In this work, we explore the excitation features of photopigments that harvest electromagnetic radiation by constructing a simple quantum-mechanical model. Our analysis suggests that the primary Earth-based photopigments for photosynthesis may not function efficiently at wavelengths $> 1.1$ $\mu$m. In the context of (hypothetical) extrasolar photopigments, we calculate the potential number of conjugated $\pi$-electrons ($N_\star$) in the relevant molecules, which can participate in the absorption of photons. By hypothesizing that the absorption maxima of photopigments are close to the peak spectral photon flux of the host star, we utilize the model to estimate $N_\star$. As per our formalism, $N_\star$ is modulated by the stellar temperature, and is conceivably higher (lower) for planets orbiting stars cooler (hotter) than the sun; exoplanets around late-type M-dwarfs might require an $N_\star$ twice that of the Earth. We conclude the analysis with a brief exposition of how our model could be empirically tested by future observations. \\
\end{abstract}

\section{Introduction}\label{SecIntro}
With the ongoing explosion in the number of exoplanets, there has been a commensurate rise in interest toward ascertaining/establishing: (1) the conditions that render a planet habitable, and (2) the traits of the putative biospheres \citep{LBC09,CBB16,Co20}. This endeavor, in addition to its theoretical import, will also have real consequences in shaping and prioritizing the selection of target planetary systems for in-depth characterization by forthcoming telescopes \citep{FAD18,SKP18}.

Although habitability and potential biospheres are clearly constrained by the properties of the world (e.g., planet or moon) in question, it is evident that these aspects are also conditioned by the host star(s). The inner and outer limits of the habitable zone (HZ)---namely, the region surrounding the star where surface temperatures on rocky planets could be conducive for liquid water---is manifestly regulated by the stellar spectral type \citep{Dole,KWR93,KRK13}.\footnote{The HZ has a labyrinthine history that stretches back to the nineteenth century and beyond \citep{LM21}.} Aside from the HZ, a multitude of vital processes on habitable planets such as abiogenesis and atmospheric retention (reviewed, for instance, in \citealt{ML19}) may be profoundly influenced by the star.

If we turn our attention to the attributes of the conjectured biospheres, photosynthesis immediately springs to mind. The vast majority of current biomass on Earth is dependent, either directly or indirectly, on (oxygenic) photosynthesis \citep{BPM18}; this is not surprising given the bountiful supply of sunlight. Since the stellar temperature of different spectral types is subjected to considerable variation, and consequently so is the ensuing spectral energy distribution, it would appear reasonable to surmise that the modes of photosynthesis might diverge from those prevalent on Earth. 

Unraveling the potential characteristics of extrasolar photosynthesis is especially valuable because it can engender distinctive gaseous (e.g., molecular oxygen) and surface (e.g., vegetation red edge) biosignatures \citep{SKP18,ML21}. Thus, acquiring a deeper knowledge about how photosynthesis may function elsewhere has direct ramifications in the quest for biosignatures. This topic has been extensively investigated in the twenty-first century \citep{WR02,KSG07,KST07,GW17,TMT17,RLR18,LCP18,LCP21,LiLo19,LL20,LL21,CAB21,CIC21}, and even in earlier publications \citep{AR73,WGP79,HDJ99}.

In this paper, we will focus on reexamining the oft-employed assumptions that the pigments facilitating the transduction of light energy (viz., photopigments) must necessarily belong to the family of chlorophylls and other such Earth-based molecules and that photosynthetically active radiation must, to a substantial extent, coincide with visible light. By computing the possible number of conjugated electrons that are involved in photosynthesis, we suggest that this quantity might be modulated by the spectral type of the star. We elucidate the theoretical model in Section \ref{SecMod} and analyze the attendant implications in Section \ref{SecDisc}.

\section{Model description}\label{SecMod}
In this Section, we describe the model used for estimating the traits of putative photopigments on temperate exoplanets around different stars.

\subsection{Model for electronic excitation}\label{SSecElecExc}
It is a well-known fact that the excitation of electrons is a vital prerequisite for photosynthesis, as expounded in \citet[Chapter 16]{LBK00}, \citet[Chapter 2]{FR07} and \citet[Chapter 1]{REB14}. A panoply of models have, therefore, sought to relate key characteristics of photopigments with their absorption spectra. A crucial parameter in these models is the number of conjugated $\pi$-electrons ($N_\star$). Since these electrons can be delocalized and are liable to excitation in response to a flux of photons \citep{MPJ16}, their number is a determinant of the pigment characteristics. 

Attempts to quantify the aforementioned relationship span a broad range of complexity: in this work, we will draw on the basic free-electron model propounded in the 1940s \citep{NSB48,HK48,HK49,WTS49}. Our rationale for doing so is twofold. First and foremost, as we shall explicate hereafter, it is endowed with a sufficient degree of accuracy and realism, while retaining simplicity and transparency. Second, as we are adopting an astrobiological perspective where we must deal with a lack of concrete data, it is desirable to keep the model generic and minimize the proliferation of free parameters that could take on arbitrary values. We will accordingly, for the most part, mirror the approach delineated in \citet[Chapter 18]{PKTG12}. 

We will model the conjugated $\pi$-electrons as though they are placed within a ``well'' of infinite depth, with the photopigment comprising the real-world counterpart of the latter. Although the derivation is fairly elementary and can be found in any standard book on quantum mechanics \citep[e.g.][Section 2.2]{GS18}, we will recapitulate the salient steps below for the benefit of the general readership. The wave function $\psi(x)$ for an electron confined within an infinite well obeys the time-independent Schr{\"o}dinger equation as follows:
\begin{equation}
    - \frac{\hbar^2}{2 m_e} \frac{d^2 \psi}{d x^2} = \mathcal{E} \psi,
\end{equation}
where $m_e$ is the mass of the electron and $\mathcal{E}$ signifies the energy. The two boundary conditions for the system are $\psi(0) = 0$ and $\psi(L) = 0$, where $L$ is the width of the infinite well. It is straightforward to verify that the mathematical solution of the above ordinary differential equation (ODE) is given by
\begin{equation}
    \psi(x) = C_1 \sin\left(k x\right) + C_2 \cos\left(k x\right),
\end{equation}
where $k = \sqrt{2 m_e \mathcal{E}/\hbar^2}$. The boundary condition at the origin demands $C_2=0$ implying $ \psi(x) = C_1 \sin\left(k x\right)$, and the boundary condition at the edge of the box leads to the quantization condition 
\begin{equation}\label{kcond}
    k L = n \pi,
\end{equation}
where $n$ is an integer. The resulting expression for the quantized energy of the $n$-th level ($\mathcal{E}_n$) is
\begin{equation}\label{EnLevel}
    \mathcal{E}_n = \frac {k^2\hbar^2}{2m_e}=\frac{n^2 h^2}{8 m_e L^2}.
\end{equation}
We will now suppose that the conjugated $\pi$-electrons are treated as effectively being situated on a chain or loop \citep{JRP49,GT92}, with a mean spacing of $\ell_\star$. By drawing on this assumption, it becomes apparent that $L = \left(N_\star - 1\right) \ell_\star$ in (\ref{EnLevel}).

We are interested in the minimal energy ($\Delta \mathcal{E}$) theoretically needed for electronic excitation and how that ties in with the longest viable photon wavelength $\lambda_\mathrm{max}$. With respect to the former, we are basically attempting to gauge the theoretical limit of the HOMO-LUMO gap \citep{AP09}, where these abbreviations stand for highest occupied molecular orbital (HOMO) and lowest unoccupied molecular orbital (LUMO). The wavelength $\lambda_\mathrm{max}$ in terms of $\Delta \mathcal{E}$ is calculated to be
\begin{equation}\label{wavecrit}
    \lambda_\mathrm{max} = \frac{h c}{\Delta \mathcal{E}}. 
\end{equation}
If we express $\Delta \mathcal{E}$ in terms of $N_\star$ and $\ell_\star$, our preceding objective is fulfilled. As indicated in the prior paragraph, we must investigate the highest occupied level. As electrons are fermions, each level contains only two of them as per the Pauli exclusion principle, with this duo differing in their spin quantum numbers. Hence, if an even number of electrons existed in total, the highest occupied level would correspond to $n = N_\star/2$, whereas for an odd number of electrons it would be $n = \left(N_\star + 1\right)/2$. If $N_\star$ is roughly an order of magnitude larger than unity, which turns out to be generally valid as illustrated hereafter, the two cases yield similar values. 

Hence, after adopting $n = N_\star/2$ and substituting this into (\ref{EnLevel}), we arrive at
\begin{equation}\label{EnHigh}
    \mathcal{E}_{N_\star/2} = \frac{h^2 N_\star^2}{32 m_e \ell_\star^2 \left(N_\star - 1\right)^2},
\end{equation}
which is the highest occupied energy level. The energy required to excite an electron from this level to the next level $n = N_\star/2 + 1$, which is unoccupied by construction, is computed by first recognizing that
\begin{equation}\label{EnUnOcc}
    \mathcal{E}_{(N_\star/2) + 1} = \frac{h^2 \left(N_\star + 2\right)^2}{32 m_e \ell_\star^2 \left(N_\star - 1\right)^2},
\end{equation}
which follows from substituting $n = N_\star/2 + 1$ in (\ref{EnLevel}). As a consequence, the theoretical minimum excitation energy $\Delta \mathcal{E}$ is therefore given by
\begin{eqnarray}\label{DeltaE}
    \Delta \mathcal{E} &=& \mathcal{E}_{(N_\star/2) + 1} -  \mathcal{E}_{N_\star/2} \nonumber \\
    &=& \frac{h^2 \left(N_\star + 1\right)}{8 m_e \ell_\star^2 \left(N_\star - 1\right)^2}.
\end{eqnarray}
It is worth recalling that $\Delta \mathcal{E}$ is the theoretical minimum photon energy needed for the requisite excitation, owing to which the ensuing wavelength represents an upper bound; to put it differently, if the wavelength is greater than $\lambda_\mathrm{max}$, then electronic excitation ought not be feasible. After substituting this expression into (\ref{wavecrit}) and solving for $\lambda_\mathrm{max}$, we duly end up with
\begin{eqnarray}\label{LamMax}
  \lambda_\mathrm{max}\, &=& \,\frac{8 m_e c}{h}\, \ell_\star^2\, \frac{\left(N_\star - 1\right)^2}{N_\star + 1} \nonumber \\
  &\approx&\, 65\,\mathrm{nm}\,\left(\frac{\ell_\star}{1.4\,\AA}\right)^2 \frac{\left(N_\star - 1\right)^2}{N_\star + 1},
\end{eqnarray}
where $\ell_\star$ has been normalized by $1.4\,\AA \equiv 0.14$ nm since it corresponds to the characteristic spacing between carbon atoms involved in conjugated bonds, in view of the empirical data adumbrated in \citet[pg. 64]{FR07} and \citet[pg. 498]{AP09}.

\begin{figure}
\includegraphics[width=7.5cm]{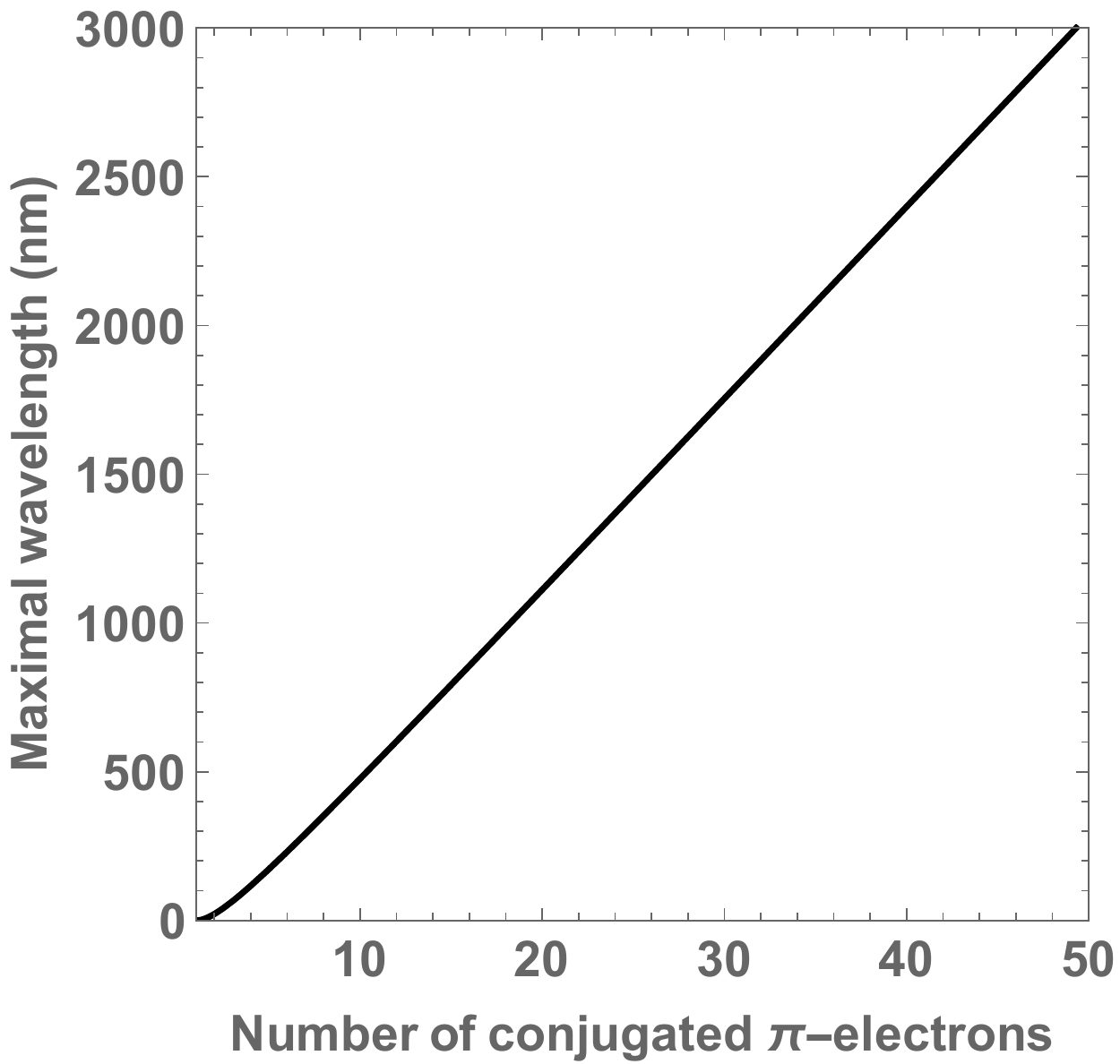} \\
\caption{The maximum wavelength that may theoretically permit electronic excitation and photosynthesis ($\lambda_\mathrm{max}$) as a function of the number of conjugated $\pi$-electrons in putative photopigments ($N_\star$) by employing (\ref{LamMax}).}
\label{FigMaxWave}
\end{figure}

In view of the simplifications invoked in deriving (\ref{LamFin}), it is natural to inquire whether this model is accurate. A bevy of studies have established that the free-electron paradigm constitutes a tenable approximation of more sophisticated treatments \citep{JRP49,WTS49,GT92}. More specifically, we note that this model has been compared against empirical data for molecules such as $\alpha$-oligothiophenes and exhibited good agreement \citep{DMS99}. Furthermore, the quantitative scaling of $\Delta \mathcal{E} \propto 1/N_\star$ expected for large $N_\star$---refer to (\ref{DeltaE})---is confirmed by experiments and density functional theory calculations for pyrroles, thiophenes, and furans \citep[e.g.,][Figure 2]{HZD03}.

At this stage, it is worth reiterating the significance of $\lambda_\mathrm{max}$; it lies in the ostensible fact that photons with $\lambda > \lambda_\mathrm{max}$ would not possess the requisite energy to excite electrons, and therefore cannot effectuate photosynthesis. We have plotted $\lambda_\mathrm{max}$ as a function of $N_\star$ in Figure \ref{FigMaxWave}.\footnote{It is reasonable to hold $\ell_\star$ fixed at its fiducial value because the mean atomic spacing is a property of the fundamental constants of nature \citep{VFW75}, and is therefore not anticipated to evince significant variation.} It is apparent from inspecting this plot that $\lambda_\mathrm{max} \propto N_\star$ for $N_\star \gg 1$ and that $\lambda_\mathrm{max} \approx 1.1$ $\mu$m for $N_\star = 20$; we shall return to this latter feature in Section \ref{SecDisc}. Conversely, if $\lambda_\mathrm{max}$ is deduced through some other empirical or theoretical method, one may potentially utilize Figure \ref{FigMaxWave} to infer $N_\star$ of the photopigments.

Until now, we have emphasized $\Delta \mathcal{E}$, which theoretically embodies the minimum energy for electronic excitation. In reality, however, this extreme limit is unlikely to suffice for excitation of electrons due to a number of intricate biochemical and biophysical processes. Laboratory assays suggest that, for the likes of chlorophylls, the excitation energy must be approximately twice that of $\Delta \mathcal{E}$ \citep{SCS75,GPRS,HIM16,BVTM}, which is verifiable after substituting $\ell_\star = 1.4\,\AA$ \citep[pg. 64]{FR07} and $N_\star = 20$ \citep[pg. 110]{RG69} in (\ref{DeltaE}). Hence, it seems credible to assume that the actual wavelength needed for robust excitation is roughly estimated by means of $\lambda \approx hc/\left(2 \Delta \mathcal{E}\right)$, which is expressible in terms of $N_\star$ as
\begin{eqnarray}\label{LamFin}
  \lambda\, &=& \,\frac{4 m_e c}{h}\, \ell_\star^2\, \frac{\left(N_\star - 1\right)^2}{N_\star + 1} \nonumber \\
  &\approx&\, 33\,\mathrm{nm}\,\left(\frac{\ell_\star}{1.4\,\AA}\right)^2 \frac{\left(N_\star - 1\right)^2}{N_\star + 1}.
\end{eqnarray}

Now, let us consider a few specific photopigments on Earth and assess the reliability of our modeling. To err on the side of caution, we restrict our attention to pigments explicitly involved in photosynthesis. For this reason, we opt not to consider microbial rhodopsins (e.g., proteorhodopsins)---which harvest as much (or more) solar energy as chlorophylls in Earth's marine environments \citep{GCR19,HTD21}---because they do not take part in photoautotrophy. The rhodopsin family, which exhibits absorption maxima of $\lesssim 550$ nm \citep{MWS03,IKN21} that are distinctly lower than those corresponding to certain (bacterio)chlorophylls, is posited to have been widespread on early Earth and similar worlds \citep{DSS21}, and might even be capable of photoautotrophy in principle \citep{LRR18}. Since there is no such concrete empirical evidence hitherto for the latter, we do not analyze this pigment hereafter as stated above; however, if and when appropriate, it is straightforward to use the fact that rhodopsin has six conjugated bonds \citep{SKB21}.

The pigment $\beta$-carotene, which is widespread in photosynthetic organisms, is endowed with $11$ conjugated bonds \citep[Figure 9.6]{RG69}. Therefore, upon specifying $N_\star = 22$ in (\ref{LamFin}), we obtain $\lambda \approx 633$ nm, which is $\lesssim 1.5$ times higher than the absorption maxima of $\beta$-carotene at $\sim 425$, $450$, and $480$ nm \citep[Table 9.2]{RG69}. Next, on turning our attention toward chlorophyll $a$, a predominant pigment in oxygenic photosynthesis \citep{JNN00}, this molecule has $10$ conjugated bonds \citep[pg. 110]{RG69}. After specifying $N_\star = 20$ in (\ref{LamFin}), we end up with $\lambda \approx 567$ nm; this value is broadly compatible with the absorption peaks at $\sim 435$ nm and $\sim 670$-$700$ nm documented for chlorophyll $a$ \citep[pg. 684]{SKP18}. 

Thus, the $\lambda$ predicted by our approach, and the wavelengths of the absorption peaks of Earth-based photopigments are manifestly not far removed from one another. For the sake of argument, suppose that certain \emph{abiotic} factors govern the wavelengths at which photopigments absorb strongly. If this wavelength regulated by such nonbiological processes (denoted by $\lambda_\mathrm{opt}$) were known, one could accordingly constrain $N_\star$ by invoking (\ref{LamFin}). The premise implicit herein is that these molecules were gradually adapted over the course of evolution such that their eventual absorption was rendered efficacious at $\lambda_\mathrm{opt}$.\footnote{To put it another way, even if photopigments that can absorb at longer or shorter wavelengths than $\lambda_\mathrm{opt}$ (akin to, say, bacteriochlorophylls) are initially favored, molecules with absorption maxima close to $\lambda_\mathrm{opt}$ might evolve over time (as outlined in Section \ref{SSecPeakAbs}), along the lines of chlorophylls on Earth.} Therefore, if we presume that this picture is tenable, it may become feasible to establish a direct connection between putative abiotic phenomena and the structure of extrasolar photopigments, which we examine further in the upcoming Sections \ref{SSecPeakAbs} and \ref{SecDisc}.

Needless to say, this potential link should not be viewed as exact because evolution is patently not a ``perfect'' optimization process; in other words, it is by no means assured that $N_\star$ for the potential photopigments would be precisely identical to what one will obtain from (\ref{LamFin}) after equating it with $\lambda_\mathrm{opt}$.

\subsection{Peak absorption wavelength of photopigments}\label{SSecPeakAbs}
The question of what wavelengths may be typically associated with the absorption peaks of photopigments on exoplanets has not been extensively investigated, barring a few notable publications \citep{KSG07,KST07,LCP18,LCP21,ML21}. One key problem is that most of these studies involve several parameters that are currently indeterminate or poorly resolved on other worlds: examples include atmospheric composition, the so-called relative cost parameter of \citet{MVG10}, and the availability of cofactors such as NADP$^+$ \citep{RM09}. Hence, we opt to construct here a simpler model with no unknown parameters; while it trades some precision, it is still fairly accurate (as expounded later). 

The essence of our working hypothesis, paralleling \citet{KSG07} and \citet[Chapter 4.3.5]{ML21}, lies in positing that the peak absorbance of photopigments occurs in the vicinity of the wavelength where the spectral photon flux ($n_\lambda$) of the host star attains its maximal value; the star is treated as a blackbody with temperature $T_\star$. We remark that this criterion is consistent with the penultimate paragraph of Section \ref{SSecElecExc} since the abiotic factor would be the stellar electromagnetic spectrum. The spectral photon flux for the star consequently is evaluated using
\begin{equation}
    n_\lambda = \frac{2 c}{\lambda^4}\left[\exp\left(\frac{h c}{\lambda k_B T_\star}\right)-1\right]^{-1}.
\end{equation}
Since $\lambda_\mathrm{opt}$ is the maximum of $n_\lambda$, it is determined by solving $d n_\lambda/d \lambda = 0$, and that simplifies to  
\begin{equation}\label{LamOpt}
\lambda_\mathrm{opt} \approx 3.67 \times 10^6\,\mathrm{nm}\,\left(\frac{T_\star}{1\,\mathrm{K}}\right)^{-1}.
\end{equation}
It is necessary to assess how realistic our model is when it comes to actually predicting the peak absorbance of photopigments, which we tackle below.

Naturally, the sun is our fiducial star---on substituting $T = T_\odot \equiv 5780$ K (viz., the temperature of the Sun), we end up with $\lambda_\mathrm{opt} \approx 635$ nm. This value constitutes a good match with the empirical absorption peaks of $\sim 630$-$700$ nm confirmed for chlorophylls $a$, $b$, and $c$ \citep[Table 1]{SKP18}. We have emphasized this trio of molecules because they are widely prevalent in oxygenic photoautotrophs. These lifeforms, in turn, comprise the predominant source of biomass on Earth \citep{BPM18}, and the evolution of oxygenic photosynthesis proved to be a major evolutionary innovation that transformed Earth's biosphere \citep{Knoll}. However, it is worth recognizing that the absorption maxima of chlorophyll $d$ and $f$ occur at slightly longer wavelengths of $710$-$740$ nm \citep{MKB13,GZR14,NMS18}.

Now, let us turn our attention to other stars, with the corresponding spectral type provided in parentheses. By applying (\ref{LamOpt}), we obtain $\lambda_\mathrm{opt} \approx 515$ nm for $T_\star = 7120$ K (F2V), $\lambda_\mathrm{opt} \approx 635$ nm for $T_\star = T_\odot$ (G2V), $\lambda_\mathrm{opt} \approx 794$ nm for $T_\star = 4620$ K (K2V), and $\lambda_\mathrm{opt} \approx 1375$ nm for $T_\star = 2670$ K (M7V). To put it another way, the absorption peak may switch to bluish wavelengths for hotter stars and near-infrared (near-IR) wavelengths for cooler stars. In order to assess how good a predictor (\ref{LamOpt}) is, we compare the formula with absorption maxima extracted from the detailed numerical model of \citet[Table 1]{LCP21}, which was itself partly predicated on \citet{MVG10}. 

The numerical model of \citet{LCP21} cited in the preceding paragraph yielded $\lambda_\mathrm{opt} \approx 468$ and $476$ nm for $T_\star = 7120$ K, $\lambda_\mathrm{opt} \approx 644$ and $672$ nm for $T_\star = 5780$ K, $\lambda_\mathrm{opt} \approx 675$, $711$, and $746$ nm for $T_\star = 4620$ K, and $\lambda_\mathrm{opt} \approx 987$ and $1050$ nm for $T_\star = 2670$ K, which clearly agrees well with our analytical model. We caution that (\ref{LamOpt}) might be somewhat inaccurate for early M-dwarfs \citep[Table 1]{LCP21}, although the potential discrepancy is diminished if we compare our calculations instead with \citet[Figure 2]{LCP18}.

As remarked a couple of paragraphs before, when one considers lower stellar temperatures, $\lambda_\mathrm{opt}$ shifts to near-IR wavelengths as per (\ref{LamOpt}), while the opposite tendency is forecast for hotter temperatures. Aside from the quantitative comparison furnished above, we note that this qualitative trend is consistent with previous studies, which have generally suggested that photosynthesis on M-dwarfs might operate at near-IR wavelengths and that the spectral edge of photosynthetic organisms may be manifested in this regime \citep{HDJ99,WR02,KST07,LL19}; however, refer to \citet{TMT17} and \citet{GW17} for contrasting takes.

\section{Discussion and Conclusions}\label{SecDisc}

\begin{figure}
\includegraphics[width=7.5cm]{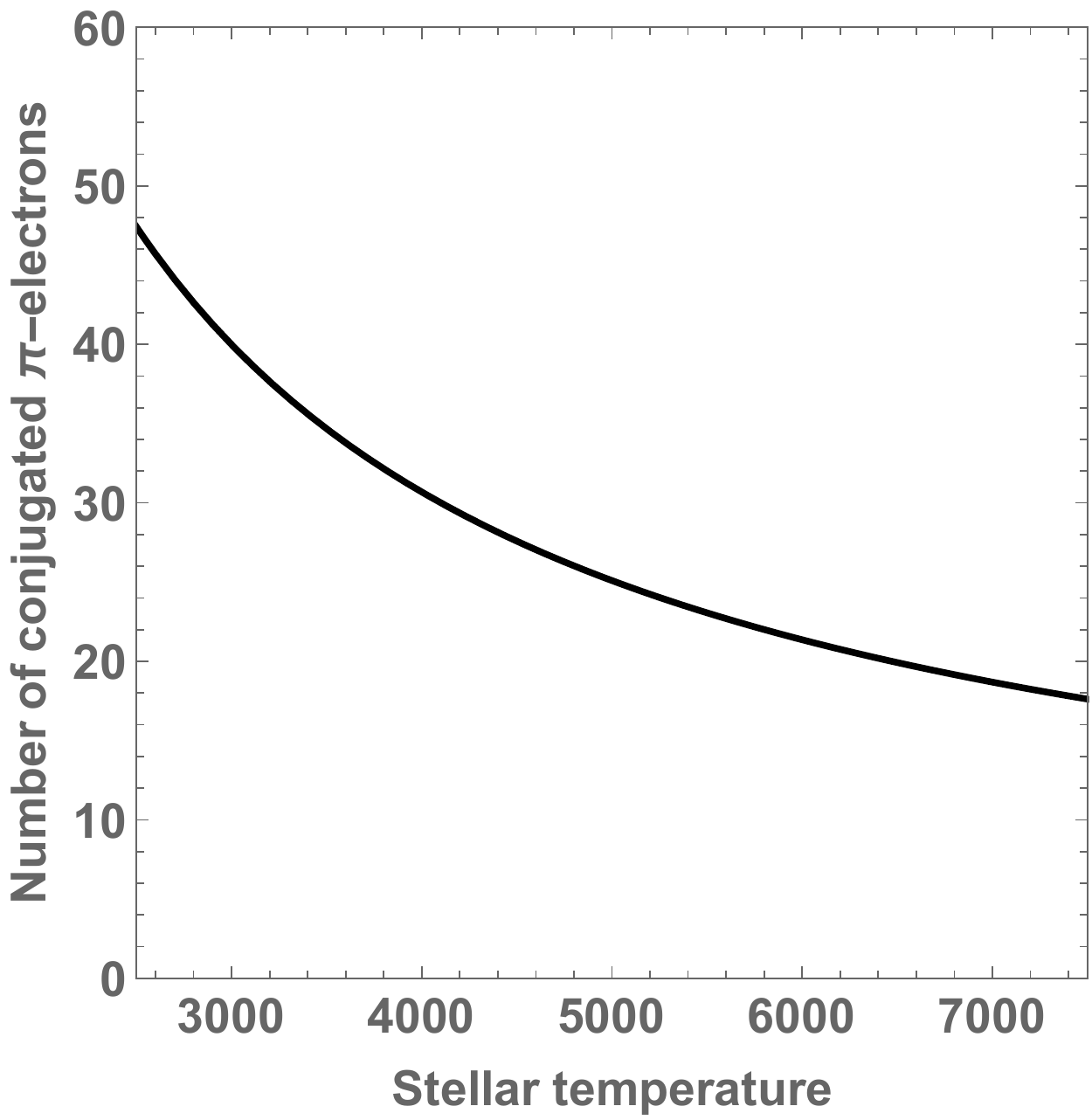} \\
\caption{The number of conjugated $\pi$-electrons in putative photopigments ($N_\star$) susceptible to excitation as a function of the stellar temperature ($T_\star$). The range of stellar temperatures broadly spans F-, G-, K-, and M-type stars.}
\label{FigPhotoElec}
\end{figure}

The first noteworthy result from our analysis is the maximal wavelength theoretically conducive to the excitation of electrons ($\lambda_\mathrm{max}$), an absolute condition for initiating photosynthesis. We determined $\lambda_\mathrm{max}$ in (\ref{LamMax}) and plotted it as a function of $N_\star$ in Figure \ref{FigMaxWave}. In view of the fact that $N_\star = 20$ or $N_\star = 22$ for the principal photopigments of interest---namely, carotenoids, chlorophylls, and bacteriochlorophylls \citep{RG69,CNG08,PKTG12}---we end up with $\lambda_\mathrm{max} \approx 1.1$-$1.25$ $\mu$m from (\ref{LamMax}).

Hence, for Earth-based photopigments, we anticipate that wavelengths greater than $\lambda_\mathrm{max}$ are ostensibly not suitable for photosynthesis. This prediction from our quantitative model displays excellent agreement with laboratory investigations, which have seemingly failed to excite electrons in primary photopigments such as chlorophylls by means of photons whose wavelengths are $> 1.1$ $\mu$m (\citealt{KSG07,KST07}; see also \citealt{LCP18}). Moreover, if extrasolar photopigments are essentially identical to those on Earth, which is, however, not necessarily the case as propounded hereafter, the same limits derived for $\lambda_\mathrm{max}$ in the prior paragraph could thus prove to be applicable. In this specific scenario, the instantiation of extraterrestrial photosynthesis at wavelengths $> 1.1$ $\mu$m may not be rendered viable.

Hitherto, we have centered our attention on $\lambda_\mathrm{max}$. We will now focus on $N_\star$, and thence explore the possibility that photopigments on other worlds might diverge from those on Earth. We can gauge $N_\star$ by substituting (\ref{LamOpt}) into (\ref{LamFin}), which is tantamount to stating that photopigments are quasi-optimized to operate at wavelengths similar to those where the spectral photon flux of the star is maximized. This procedure generates a quadratic equation for $N_\star$ in terms of $T_\star$ and $\ell_\star$:
\begin{equation}\label{Nstarint}
 1.1 \times 10^5 \left(\frac{T_\star}{1\,\mathrm{K}}\right)^{-1} \approx \left(\frac{\ell_\star}{1.4\,\AA}\right)^2 \frac{\left(N_\star - 1\right)^2}{N_\star + 1}.
\end{equation}
As per our previous exposition, $N_\star \gg 1$ is expected to be quite valid, which allows us to simplify the above equation and thereupon obtain
\begin{equation}\label{Nstarfin}
    N_\star \approx 1.1 \times 10^5\, \left(\frac{T_\star}{1\,\mathrm{K}}\right)^{-1} \left(\frac{\ell_\star}{1.4\,\AA}\right)^{-2}.
\end{equation}
Let us specify $T_\star = T_\odot$ in (\ref{Nstarfin}) and hold $\ell_\star$ fixed at its fiducial value, which accordingly yields $N_\star \approx 19$. This estimate is evidently proximate to the number of conjugate electrons in chlorophyll ($N_\star = 20$) and $\beta$-carotene ($N_\star = 22$) \citep[pg. 736]{PKTG12}, thereby lending credence to our analysis and hypothesis; the \emph{a priori} requirement of $N_\star \gg 1$ is also satisfied. On solving the full quadratic equation for $N_\star$ given by (\ref{Nstarint}), we arrive at $N_\star \approx 22$, which obviously exhibits excellent agreement with the aforementioned molecules.

In Figure \ref{FigPhotoElec}, we have depicted the variation of $N_\star$ with the stellar temperature, whose range encompasses F-, G-, K-, and M-type stars, by employing (\ref{Nstarint}). The first major takeaway from Figure \ref{FigPhotoElec} is that $N_\star$ is a monotonically decreasing function of $T_\star$. The second, and arguably more crucial, feature is that $N_\star$ becomes noticeably high for M-dwarfs, especially for late M-dwarfs. For instance, if we consider the famous TRAPPIST-1 planetary system with $T_\star \approx 2566$ K \citep[Table 7]{ADG21}, we duly obtain $N_\star \approx 46$ (for $\ell_\star \approx 1.4\,\AA$), which is about twice that of chlorophyll $a$ and $\beta$-carotene.

On the basis of these results, we might potentially draw a tentative conclusion. If we posit that our hypothesis is tenable, to wit, that $N_\star$ is substantively constrained by the peak absorbance of photopigments---which is presumably close to the peak of the spectral photon flux, as illustrated in Section \ref{SSecPeakAbs}---it is not implausible that, if planets around M-dwarfs harbor photopigments, their structures would be very different from those of the widespread chlorophylls. The latter are derived from porphine (and its variants), which is endowed with $11$ conjugated bonds ($N_\star = 22$), as seen from consulting \citet[Figure 9.2]{RG69}. 

Hence, this value contrasts with $N_\star$ determined for late M-dwarfs via Figure \ref{FigPhotoElec}, which can become almost twice as large compared to the estimate for the Sun when all other factors are held fixed. To put it another way, as per our model, the photopigments on these worlds cannot be simply categorized as canonical chlorophylls and carotenes, both of which are characterized by $N_\star \approx 20$. We will not speculate herein about what their structures may resemble, owing to the paucity of concrete data in this regard. It might be possible for the putative photopigments to comprise two porphine molecules linked together in some fashion, which could yield the desired number of conjugated $\pi$-electrons.

In contrast, a few notable proposals have advocated that chlorophyll constitutes a ``universal pigment'' by virtue of its advantages vis-\`a-vis stability and structure \citep{GW59,GW74,RW04}. Needless to say, if this school of thought were correct, it would suppress the prospects for novel pigments along the lines intimated above. In this scenario, even on planets orbiting M-dwarfs, one would expect to find chlorophylls and their absorption peaks will occur at $\lesssim 750$ nm if the Earth does represent a suitable benchmark. However, on account of the relative scarcity of such photons, the net primary productivity of these biospheres would be potentially much lower than that of Earth \citep{WGP79,WR02,KST07,LCP18,LiLo19,LL20,LL21}.

To summarize, there exist a couple of distinct outcomes that merit consideration. On the one hand, as suggested in this paper, extrasolar photopigments may not belong to the group of chlorophylls and these putative molecules could instantiate strong absorption in the near-infrared on M-dwarf exoplanets, which we choose to label hypothesis \#1. On the other hand, it is conceivable that chlorophylls are truly quasi-universal in nature and that the absorption peaks of photopigments on planets around all types of stars might therefore transpire at $\lesssim 750$ nm. Fortunately, however, it ought to be straightforward to empirically test/falsify our hypothesis by future observations as delineated below. 

If the spectral edge (i.e., documented steep change in reflectance) associated with photopigments is nearly independent of the stellar spectral type and transpires in the vicinity of red wavelengths \citep{STSF}, data from the reflected light of exoplanets (hosting such biomolecules) garnered by future direct-imaging surveys \citep{FAD18} may effectively suffice to eliminate our model. In contrast, if the spectral edge is manifested in accordance with (\ref{LamOpt}), it could lend credence to our model. However, other paradigms---for example, linking photosystems in series, thereby amounting to multiple Z-schemes operating in tandem \citep{KST07}, which we christen hypothesis \#2---might shift the spectral edge to longer wavelengths on cooler worlds. 

It is essential to recognize that these spectral edges are more prominent and detectable for land-based photosynthetic organisms such as embryophytes on Earth \citep{STSF,CKR09}. Hence, the ensuing discussion is likely of diminished applicability to ocean worlds (sans landmasses)---which are anticipated to be quite prevalent on the basis of exoplanet surveys \citep{ZJS19,MDA20}---since the signatures may not be as pronounced \citep[e.g.,][Figure 10]{SKP18} and water additionally modulates the spectral niches of photosynthesis \citep{SHS07,HHS21}; see, however, the simulations by \citet{OMK18,OMK19}.\footnote{Moreover, certain characteristics of aquatic photosynthesis (e.g., euphotic zone depth and net primary productivity) are sensitive to the spectral type of the star \citep[e.g.,][]{LL20,LL21}.}

Thus, coupled photosystems predicated purely on Earth-based photopigments (hypothesis \#2) may engender near-IR spectral edges on M-dwarf exoplanets, perhaps obviating the necessity for ``exotic'' photopigments (hypothesis \#1) of the kind theorized in this work. It is consequently imperative to come up with diagnostics for differentiating between these two hypotheses. In this context, we highlight that spectral edges at longer wavelengths associated with hypothesis \#2 are near-integer multiples of $350$ nm \citep[pg. 539]{WR02}, whereas this discrete pattern is not expected for hypothesis \#1. Hence, the location of the spectral edge on various worlds could aid in distinguishing between hypotheses \#1 and \#2. Furthermore, hypothesis \#1 allows for shorter wavelengths on hotter worlds, while this does not seem tenable for hypothesis \#2.

In view of the significance of photosynthesis on Earth---from both the evolutionary and ecological standpoints \citep{FR07,REB14,Knoll}---as well as the fact that it can generate powerful gaseous and surface biosignatures that permit the detection of extraterrestrial life \citep{SKP18,ML21}, the importance of gauging the electronic excitation properties of photopigments is readily apparent. A simple theoretical model with a minimal set of unknown parameters, which is proposed, solved, and analyzed in this paper, gives qualitative and quantitative predictions pertaining to the possible relationship between the spectral type of the host star and the nature of suitable biomolecules (analogs of terrestrial chlorophylls) that harness starlight to create biomass. We hope that some parts of this paper will be relevant to the understanding and interpretation of the collective, fast-increasing knowledge of exoplanets.

\acknowledgments
A.B. acknowledges support from the Italian Space Agency (ASI, DC-VUM-2017-034, grant number 2019-3 U.O Life in Space) and by the grants FQXi-MGA-1801 and FQXi-MGB-1924 from the Foundational Questions Institute and Fetzer Franklin Fund, a donor advised fund of Silicon Valley Community Foundation.

\bibliographystyle{aasjournal}
\bibliography{PhotoPigment}

\begin{thebibliography}{}
\expandafter\ifx\csname natexlab\endcsname\relax\def\natexlab#1{#1}\fi
\providecommand{\url}[1]{\href{#1}{#1}}
\providecommand{\dodoi}[1]{doi:~\href{http://doi.org/#1}{\nolinkurl{#1}}}
\providecommand{\doeprint}[1]{\href{http://ascl.net/#1}{\nolinkurl{http://ascl.net/#1}}}
\providecommand{\doarXiv}[1]{\href{https://arxiv.org/abs/#1}{\nolinkurl{https://arxiv.org/abs/#1}}}

\bibitem[{{Agol} {et~al.}(2021){Agol}, {Dorn}, {Grimm}, {Turbet}, {Ducrot},
  {Delrez}, {Gillon}, {Demory}, {Burdanov}, {Barkaoui}, {Benkhaldoun},
  {Bolmont}, {Burgasser}, {Carey}, {de Wit}, {Fabrycky}, {Foreman-Mackey},
  {Haldemann}, {Hernandez}, {Ingalls}, {Jehin}, {Langford}, {Leconte},
  {Lederer}, {Luger}, {Malhotra}, {Meadows}, {Morris}, {Pozuelos}, {Queloz},
  {Raymond}, {Selsis}, {Sestovic}, {Triaud}, \& {Van Grootel}}]{ADG21}
{Agol}, E., {Dorn}, C., {Grimm}, S.~L., {et~al.} 2021, Planet. Sci. J., 2, 1,
  \dodoi{10.3847/PSJ/abd022}

\bibitem[{{Atkins} \& {de Paula}(2009)}]{AP09}
{Atkins}, P., \& {de Paula}, J. 2009, {Elements of Physical Chemistry}, 5th
  edn. (W. H. Freeman \& Co., New York)

\bibitem[{{Bar-On} {et~al.}(2018){Bar-On}, {Phillips}, \& {Milo}}]{BPM18}
{Bar-On}, Y.~M., {Phillips}, R., \& {Milo}, R. 2018, Proc. Natl. Acad. Sci.
  USA, 115, 6506, \dodoi{10.1073/pnas.1711842115}

\bibitem[{{Bayliss}(1948)}]{NSB48}
{Bayliss}, N.~S. 1948, J. Chem. Phys., 16, 287, \dodoi{10.1063/1.1746869}

\bibitem[{{Blankenship}(2014)}]{REB14}
{Blankenship}, R.~E. 2014, {Molecular Mechanisms of Photosynthesis}, 2nd edn.
  (Wiley-Blackwell, Chichester)

\bibitem[{{Buscemi} {et~al.}(2021){Buscemi}, {Vona}, {Trotta}, {Milano}, \&
  {Farinola}}]{BVTM}
{Buscemi}, G., {Vona}, D., {Trotta}, M., {Milano}, F., \& {Farinola}, G.~M.
  2021, Adv. Mater. Technol., 2100245, \dodoi{10.1002/admt.202100245}

\bibitem[{{Claudi} {et~al.}(2021){Claudi}, {Alei}, {Battistuzzi}, {Cocola},
  {Erculiani}, {Pozzer}, {Salasnich}, {Simionato}, {Squicciarini}, {Poletto},
  \& {La Rocca}}]{CAB21}
{Claudi}, R., {Alei}, E., {Battistuzzi}, M., {et~al.} 2021, Life, 11, 10,
  \dodoi{10.3390/life11010010}

\bibitem[{{Cockell}(2020)}]{Co20}
{Cockell}, C.~S. 2020, {Astrobiology: Understanding Life in the Universe}, 2nd
  edn. (John Wiley \& Sons, Hoboken)

\bibitem[{{Cockell} {et~al.}(2009){Cockell}, {Kaltenegger}, \& {Raven}}]{CKR09}
{Cockell}, C.~S., {Kaltenegger}, L., \& {Raven}, J.~A. 2009, Astrobiology, 9,
  623, \dodoi{10.1089/ast.2008.0273}

\bibitem[{{Cockell} {et~al.}(2016){Cockell}, {Bush}, {Bryce}, {Direito},
  {Fox-Powell}, {Harrison}, {Lammer}, {Landenmark}, {Martin-Torres},
  {Nicholson}, {Noack}, {O'Malley-James}, {Payler}, {Rushby}, {Samuels},
  {Schwendner}, {Wadsworth}, \& {Zorzano}}]{CBB16}
{Cockell}, C.~S., {Bush}, T., {Bryce}, C., {et~al.} 2016, Astrobiology, 16, 89,
  \dodoi{10.1089/ast.2015.1295}

\bibitem[{{Cong} {et~al.}(2008){Cong}, {Niedzwiedzki}, {Gibson}, {LaFountain},
  {Kelsh}, {Gardiner}, {Cogdell}, \& {Frank}}]{CNG08}
{Cong}, H., {Niedzwiedzki}, D.~M., {Gibson}, G.~N., {et~al.} 2008, J. Phys.
  Chem. B, 112, 10689, \dodoi{10.1021/jp711946w}

\bibitem[{{Covone} {et~al.}(2021){Covone}, {Ienco}, {Cacciapuoti}, \&
  {Inno}}]{CIC21}
{Covone}, G., {Ienco}, R.~M., {Cacciapuoti}, L., \& {Inno}, L. 2021, Mon. Not.
  R. Astron. Soc., 505, 3329, \dodoi{10.1093/mnras/stab1357}

\bibitem[{{DasSarma} \& {Schwieterman}(2021)}]{DSS21}
{DasSarma}, S., \& {Schwieterman}, E.~W. 2021, Int. J. Astrobiol., 20, 241,
  \dodoi{10.1017/S1473550418000423}

\bibitem[{{de Melo} {et~al.}(1999){de Melo}, {Silva}, {Arnaut}, \&
  {Becker}}]{DMS99}
{de Melo}, J.~S., {Silva}, L.~M., {Arnaut}, L.~G., \& {Becker}, R.~S. 1999, J.
  Chem. Phys., 111, 5427, \dodoi{10.1063/1.479825}

\bibitem[{{Dole}(1964)}]{Dole}
{Dole}, S.~H. 1964, {Habitable planets for man} (Blaisdell Pub.~Co., New York)

\bibitem[{{Falkowski} \& {Raven}(2007)}]{FR07}
{Falkowski}, P.~G., \& {Raven}, J.~A. 2007, {Aquatic photosynthesis}, 2nd edn.
  (Princeton University Press, Princeton)

\bibitem[{{Fujii} {et~al.}(2018){Fujii}, {Angerhausen}, {Deitrick},
  {Domagal-Goldman}, {Grenfell}, {Hori}, {Kane}, {Pall{\'e}}, {Rauer},
  {Siegler}, {Stapelfeldt}, \& {Stevenson}}]{FAD18}
{Fujii}, Y., {Angerhausen}, D., {Deitrick}, R., {et~al.} 2018, Astrobiology,
  18, 739, \dodoi{10.1089/ast.2017.1733}

\bibitem[{{Gale} \& {Wandel}(2017)}]{GW17}
{Gale}, J., \& {Wandel}, A. 2017, Int. J. Astrobiol., 16, 1,
  \dodoi{10.1017/S1473550415000440}

\bibitem[{{Gan} {et~al.}(2014){Gan}, {Zhang}, {Rockwell}, {Martin}, {Lagarias},
  \& {Bryant}}]{GZR14}
{Gan}, F., {Zhang}, S., {Rockwell}, N.~C., {et~al.} 2014, Science, 345, 1312,
  \dodoi{10.1126/science.1256963}

\bibitem[{{G{\'o}mez-Consarnau} {et~al.}(2019){G{\'o}mez-Consarnau}, {Raven},
  {Levine}, {Cutter}, {Wang}, {Seegers}, {Ar{\'\i}stegui}, {Fuhrman}, {Gasol},
  \& {Sa{\~n}udo-Wilhelmy}}]{GCR19}
{G{\'o}mez-Consarnau}, L., {Raven}, J.~A., {Levine}, N.~M., {et~al.} 2019, Sci.
  Adv., 5, eaaw8855, \dodoi{10.1126/sciadv.aaw8855}

\bibitem[{{Griffiths} \& {Schroeter}(2018)}]{GS18}
{Griffiths}, D.~J., \& {Schroeter}, D.~F. 2018, {Introduction to Quantum
  Mechanics}, 3rd edn. (Cambridge University Press, Cambridge)

\bibitem[{{Grimm} {et~al.}(2006){Grimm}, {Porra}, {R{\"u}diger}, \&
  {Scheer}}]{GPRS}
{Grimm}, B., {Porra}, R.~J., {R{\"u}diger}, W., \& {Scheer}, H., eds. 2006,
  {Chlorophylls and Bacteriochlorophylls: Biochemistry, Biophysics, Functions
  and Applications} (Springer, Dordrecht), \dodoi{10.1007/1-4020-4516-6}

\bibitem[{{Hassanzadeh} {et~al.}(2021){Hassanzadeh}, {Thomson}, {Deans},
  {Wenley}, {Lockwood}, {Currie}, {Morales}, {Steindler},
  {Sa{\~n}udo-Wilhelmy}, {Baltar}, \& {G{\'o}mez-Consarnau}}]{HTD21}
{Hassanzadeh}, B., {Thomson}, B., {Deans}, F., {et~al.} 2021, Environ.
  Microbiol. Rep., 13, 401, \dodoi{10.1111/1758-2229.12948}

\bibitem[{{Heath} {et~al.}(1999){Heath}, {Doyle}, {Joshi}, \&
  {Haberle}}]{HDJ99}
{Heath}, M.~J., {Doyle}, L.~R., {Joshi}, M.~M., \& {Haberle}, R.~M. 1999, Orig.
  Life Evol. Biosph., 29, 405, \dodoi{10.1023/A:1006596718708}

\bibitem[{{Hedayatifar} {et~al.}(2016){Hedayatifar}, {Irani}, {Mazarei},
  {Rasti}, {Azar}, {Rezakhani}, {Mashaghi}, {Shayeganfar}, {Anvari}, {Heydari},
  {Tabar}, {Nafari}, {Vesaghi}, {Asgari}, \& {Rahimi Tabar}}]{HIM16}
{Hedayatifar}, L., {Irani}, E., {Mazarei}, M., {et~al.} 2016, RSC Adv., 6,
  109778, \dodoi{10.1039/C6RA20226H}

\bibitem[{{Holtrop} {et~al.}(2021){Holtrop}, {Huisman}, {Stomp}, {Biersteker},
  {Aerts}, {Gr{\'e}bert}, {Partensky}, {Garczarek}, \& {van der Woerd}}]{HHS21}
{Holtrop}, T., {Huisman}, J., {Stomp}, M., {et~al.} 2021, Nat. Ecol. Evol., 5,
  55, \dodoi{10.1038/s41559-020-01330-x}

\bibitem[{{Hutchison} {et~al.}(2003){Hutchison}, {Zhao}, {Delley}, {Freeman},
  {Ratner}, \& {Marks}}]{HZD03}
{Hutchison}, G.~R., {Zhao}, Y.-J., {Delley}, B., {et~al.} 2003, Phys. Rev. B,
  68, 035204, \dodoi{10.1103/PhysRevB.68.035204}

\bibitem[{{Inoue} {et~al.}(2021){Inoue}, {Karasuyama}, {Nakamura}, {Konno},
  {Yamada}, {Mannen}, {Nagata}, {Inatsu}, {Yawo}, {Yura}, {B{\'e}ja},
  {Kandori}, \& {Takeuchi}}]{IKN21}
{Inoue}, K., {Karasuyama}, M., {Nakamura}, R., {et~al.} 2021, Commun. Biol., 4,
  362, \dodoi{10.1038/s42003-021-01878-9}

\bibitem[{{Johnson}(2016)}]{MPJ16}
{Johnson}, M.~P. 2016, Essays Biochem., 60, 255, \dodoi{10.1042/EBC20160016}

\bibitem[{{Kasting} {et~al.}(1993){Kasting}, {Whitmire}, \& {Reynolds}}]{KWR93}
{Kasting}, J.~F., {Whitmire}, D.~P., \& {Reynolds}, R.~T. 1993, Icarus, 101,
  108, \dodoi{10.1006/icar.1993.1010}

\bibitem[{{Kiang} {et~al.}(2007{\natexlab{a}}){Kiang}, {Siefert}, {Govindjee},
  \& {Blankenship}}]{KSG07}
{Kiang}, N.~Y., {Siefert}, J., {Govindjee}, \& {Blankenship}, R.~E.
  2007{\natexlab{a}}, Astrobiology, 7, 222, \dodoi{10.1089/ast.2006.0105}

\bibitem[{{Kiang} {et~al.}(2007{\natexlab{b}}){Kiang}, {Segura}, {Tinetti},
  {Govindjee}, {Blankenship}, {Cohen}, {Siefert}, {Crisp}, \&
  {Meadows}}]{KST07}
{Kiang}, N.~Y., {Segura}, A., {Tinetti}, G., {et~al.} 2007{\natexlab{b}},
  Astrobiology, 7, 252, \dodoi{10.1089/ast.2006.0108}

\bibitem[{{Knoll}(2015)}]{Knoll}
{Knoll}, A.~H. 2015, {Life on a Young Planet: The First Three Billion Years of
  Evolution on Earth}, Princeton Science Library (Princeton University Press,
  Princeton)

\bibitem[{{Kopparapu} {et~al.}(2013){Kopparapu}, {Ramirez}, {Kasting}, {Eymet},
  {Robinson}, {Mahadevan}, {Terrien}, {Domagal-Goldman}, {Meadows}, \&
  {Deshpande}}]{KRK13}
{Kopparapu}, R.~K., {Ramirez}, R., {Kasting}, J.~F., {et~al.} 2013, Astrophys.
  J., 765, 131, \dodoi{10.1088/0004-637X/765/2/131}

\bibitem[{{Kuhn}(1948)}]{HK48}
{Kuhn}, H. 1948, J. Chem. Phys., 16, 840, \dodoi{10.1063/1.1747011}

\bibitem[{{Kuhn}(1949)}]{HK49}
---. 1949, J. Chem. Phys., 17, 1198, \dodoi{10.1063/1.1747143}

\bibitem[{{Lammer} {et~al.}(2009){Lammer}, {Bredeh{\"o}ft}, {Coustenis},
  {Khodachenko}, {Kaltenegger}, {Grasset}, {Prieur}, {Raulin}, {Ehrenfreund},
  {Yamauchi}, {Wahlund}, {Grie{\ss}meier}, {Stangl}, {Cockell}, {Kulikov},
  {Grenfell}, \& {Rauer}}]{LBC09}
{Lammer}, H., {Bredeh{\"o}ft}, J.~H., {Coustenis}, A., {et~al.} 2009, Astron.
  Astrophys. Rev., 17, 181, \dodoi{10.1007/s00159-009-0019-z}

\bibitem[{{Larkum} {et~al.}(2018){Larkum}, {Ritchie}, \& {Raven}}]{LRR18}
{Larkum}, A. W.~D., {Ritchie}, R.~J., \& {Raven}, J.~A. 2018, Photosynthetica,
  56, 11, \dodoi{10.1007/s11099-018-0792-x}

\bibitem[{{Lehmer} {et~al.}(2018){Lehmer}, {Catling}, {Parenteau}, \&
  {Hoehler}}]{LCP18}
{Lehmer}, O.~R., {Catling}, D.~C., {Parenteau}, M.~N., \& {Hoehler}, T.~M.
  2018, Astrophys. J., 859, 171, \dodoi{10.3847/1538-4357/aac104}

\bibitem[{{Lehmer} {et~al.}(2021){Lehmer}, {Catling}, {Parenteau}, {Kiang}, \&
  {Hoehler}}]{LCP21}
{Lehmer}, O.~R., {Catling}, D.~C., {Parenteau}, M.~N., {Kiang}, N.~Y., \&
  {Hoehler}, T.~M. 2021, Front. Astron. Space Sci., 8, 689441,
  \dodoi{10.3389/fspas.2021.689441}

\bibitem[{{Lingam}(2021)}]{LM21}
{Lingam}, M. 2021, Int. J. Astrobiol., 20, 332,
  \dodoi{10.1017/S1473550421000203}

\bibitem[{{Lingam} \& {Loeb}(2019{\natexlab{a}})}]{ML19}
{Lingam}, M., \& {Loeb}, A. 2019{\natexlab{a}}, Rev. Mod. Phys., 91, 021002,
  \dodoi{10.1103/RevModPhys.91.021002}

\bibitem[{{Lingam} \& {Loeb}(2019{\natexlab{b}})}]{LiLo19}
---. 2019{\natexlab{b}}, Mon. Not. R. Astron. Soc., 485, 5924,
  \dodoi{10.1093/mnras/stz847}

\bibitem[{{Lingam} \& {Loeb}(2019{\natexlab{c}})}]{LL19}
---. 2019{\natexlab{c}}, Astrophys. J., 883, 143,
  \dodoi{10.3847/1538-4357/ab3f35}

\bibitem[{{Lingam} \& {Loeb}(2020)}]{LL20}
---. 2020, Astrophys. J. Lett., 889, L15, \dodoi{10.3847/2041-8213/ab6a14}

\bibitem[{{Lingam} \& {Loeb}(2021{\natexlab{a}})}]{ML21}
---. 2021{\natexlab{a}}, {Life in the Cosmos: From Biosignatures to
  Technosignatures} (Harvard University Press, Cambridge).
\newblock
  \url{https://books.google.com/books/about/Life_in_the_Cosmos.html?id=m4vpzQEACAAJ}

\bibitem[{{Lingam} \& {Loeb}(2021{\natexlab{b}})}]{LL21}
---. 2021{\natexlab{b}}, Mon. Not. R. Astron. Soc., 503, 3434,
  \dodoi{10.1093/mnras/stab611}

\bibitem[{{Lodish} {et~al.}(2000){Lodish}, {Berk}, {Zipursky}, {Matsudaira},
  {Baltimore}, \& {Darnell}}]{LBK00}
{Lodish}, H., {Berk}, A., {Zipursky}, S.~L., {et~al.} 2000, {Molecular Cell
  Biology}, 4th edn. (W. H. Freeman \& Co., New York)

\bibitem[{{Man} {et~al.}(2003){Man}, {Wang}, {Sabehi}, {Aravind}, {Post},
  {Massana}, {Spudich}, {Spudich}, \& {B{\'e}ja}}]{MWS03}
{Man}, D., {Wang}, W., {Sabehi}, G., {et~al.} 2003, EMBO J., 22, 1725,
  \dodoi{10.1093/emboj/cdg183}

\bibitem[{{Marosv{\"o}lgyi} \& {van Gorkom}(2010)}]{MVG10}
{Marosv{\"o}lgyi}, M.~A., \& {van Gorkom}, H.~J. 2010, Photosynth. Res., 103,
  105, \dodoi{10.1007/s11120-009-9522-3}

\bibitem[{{Mielke} {et~al.}(2013){Mielke}, {Kiang}, {Blankenship}, \&
  {Mauzerall}}]{MKB13}
{Mielke}, S.~P., {Kiang}, N.~Y., {Blankenship}, R.~E., \& {Mauzerall}, D. 2013,
  Biochim. Biophys. Acta Bioenerg., 1827, 255,
  \dodoi{10.1016/j.bbabio.2012.11.002}

\bibitem[{{Milo}(2009)}]{RM09}
{Milo}, R. 2009, Photosynth. Res., 101, 59, \dodoi{10.1007/s11120-009-9465-8}

\bibitem[{{Mousis} {et~al.}(2020){Mousis}, {Deleuil}, {Aguichine}, {Marcq},
  {Naar}, {Aguirre}, {Brugger}, \& {Gon{\c{c}}alves}}]{MDA20}
{Mousis}, O., {Deleuil}, M., {Aguichine}, A., {et~al.} 2020, Astrophys. J.
  Lett., 896, L22, \dodoi{10.3847/2041-8213/ab9530}

\bibitem[{{Nishio}(2000)}]{JNN00}
{Nishio}, J.~N. 2000, Plant Cell Environ., 23, 539,
  \dodoi{10.1046/j.1365-3040.2000.00563.x}

\bibitem[{{N{\"u}rnberg} {et~al.}(2018){N{\"u}rnberg}, {Morton},
  {Santabarbara}, {Telfer}, {Joliot}, {Antonaru}, {Ruban}, {Cardona}, {Krausz},
  {Boussac}, {Fantuzzi}, \& {Rutherford}}]{NMS18}
{N{\"u}rnberg}, D.~J., {Morton}, J., {Santabarbara}, S., {et~al.} 2018,
  Science, 360, 1210, \dodoi{10.1126/science.aar8313}

\bibitem[{{O'Malley-James} \& {Kaltenegger}(2018)}]{OMK18}
{O'Malley-James}, J.~T., \& {Kaltenegger}, L. 2018, Astrobiology, 18, 1123,
  \dodoi{10.1089/ast.2017.1798}

\bibitem[{{O'Malley-James} \& {Kaltenegger}(2019)}]{OMK19}
---. 2019, Astrophys. J. Lett., 879, L20, \dodoi{10.3847/2041-8213/ab2769}

\bibitem[{{Phillips} {et~al.}(2012){Phillips}, {Kondev}, {Theriot}, \&
  {Garcia}}]{PKTG12}
{Phillips}, R., {Kondev}, J., {Theriot}, J., \& {Garcia}, H.~G. 2012, {Physical
  Biology of the Cell}, 2nd edn. (Garland Science, New York)

\bibitem[{{Platt}(1949)}]{JRP49}
{Platt}, J.~R. 1949, J. Chem. Phys., 17, 484, \dodoi{10.1063/1.1747293}

\bibitem[{{Pollard}(1979)}]{WGP79}
{Pollard}, W.~G. 1979, Am. Sci., 67, 653

\bibitem[{{Rabinowitch} \& {Govindjee}(1969)}]{RG69}
{Rabinowitch}, E., \& {Govindjee}. 1969, {Photosynthesis} (John Wiley \& Sons,
  New York)

\bibitem[{{Raven} \& {Wolstencroft}(2004)}]{RW04}
{Raven}, J.~A., \& {Wolstencroft}, R.~D. 2004, in Bioastronomy 2002: Life Among
  the Stars, ed. R.~{Norris} \& F.~{Stootman}, Vol. 213 (Astronomical Society
  of the Pacific, San Francisco), 305--308

\bibitem[{{Ritchie} {et~al.}(2018){Ritchie}, {Larkum}, \& {Ribas}}]{RLR18}
{Ritchie}, R.~J., {Larkum}, A. W.~D., \& {Ribas}, I. 2018, Int. J. Astrobiol.,
  17, 147, \dodoi{10.1017/S1473550417000167}

\bibitem[{{Rueda}(1973)}]{AR73}
{Rueda}, A. 1973, Space Life Sci., 4, 469, \dodoi{10.1007/BF00930358}

\bibitem[{{Schwieterman} {et~al.}(2018){Schwieterman}, {Kiang}, {Parenteau},
  {Harman}, {DasSarma}, {Fisher}, {Arney}, {Hartnett}, {Reinhard}, {Olson},
  {Meadows}, {Cockell}, {Walker}, {Grenfell}, {Hegde}, {Rugheimer}, {Hu}, \&
  {Lyons}}]{SKP18}
{Schwieterman}, E.~W., {Kiang}, N.~Y., {Parenteau}, M.~N., {et~al.} 2018,
  Astrobiology, 18, 663, \dodoi{10.1089/ast.2017.1729}

\bibitem[{{Seager} {et~al.}(2005){Seager}, {Turner}, {Schafer}, \&
  {Ford}}]{STSF}
{Seager}, S., {Turner}, E.~L., {Schafer}, J., \& {Ford}, E.~B. 2005,
  Astrobiology, 5, 372, \dodoi{10.1089/ast.2005.5.372}

\bibitem[{{Sen} {et~al.}(2021){Sen}, {Kar}, {Borin}, \& {Schapiro}}]{SKB21}
{Sen}, S., {Kar}, R.~K., {Borin}, V.~A., \& {Schapiro}, I. 2021, Wiley
  Interdiscip. Rev. Comput. Mol. Sci., e1562, \dodoi{10.1002/wcms.1562}

\bibitem[{{Serlin} {et~al.}(1975){Serlin}, {Chow}, \& {Strouse}}]{SCS75}
{Serlin}, R., {Chow}, H.-C., \& {Strouse}, C.~E. 1975, J. Am. Chem. Soc., 97,
  7237, \dodoi{10.1021/ja00858a007}

\bibitem[{{Simpson}(1949)}]{WTS49}
{Simpson}, W.~T. 1949, J. Chem. Phys., 17, 1218, \dodoi{10.1063/1.1747145}

\bibitem[{{Stomp} {et~al.}(2007){Stomp}, {Huisman}, {Stal}, \&
  {Matthijs}}]{SHS07}
{Stomp}, M., {Huisman}, J., {Stal}, L.~J., \& {Matthijs}, H. C.~P. 2007, ISME
  J., 1, 271, \dodoi{10.1038/ismej.2007.59}

\bibitem[{{Takizawa} {et~al.}(2017){Takizawa}, {Minagawa}, {Tamura},
  {Kusakabe}, \& {Narita}}]{TMT17}
{Takizawa}, K., {Minagawa}, J., {Tamura}, M., {Kusakabe}, N., \& {Narita}, N.
  2017, Sci. Rep., 7, 7561, \dodoi{10.1038/s41598-017-07948-5}

\bibitem[{{Taubmann}(1992)}]{GT92}
{Taubmann}, G. 1992, J. Chem. Educ., 69, 96, \dodoi{10.1021/ed069p96}

\bibitem[{{Wald}(1959)}]{GW59}
{Wald}, G. 1959, Sci. Am., 201, 92, \dodoi{10.1038/scientificamerican1059-92}

\bibitem[{{Wald}(1974)}]{GW74}
---. 1974, Orig. Life, 5, 7, \dodoi{10.1007/BF00927010}

\bibitem[{{Weisskopf}(1975)}]{VFW75}
{Weisskopf}, V.~F. 1975, Science, 187, 605,
  \dodoi{10.1126/science.187.4177.605}

\bibitem[{{Wolstencroft} \& {Raven}(2002)}]{WR02}
{Wolstencroft}, R.~D., \& {Raven}, J.~A. 2002, Icarus, 157, 535,
  \dodoi{10.1006/icar.2002.6854}

\bibitem[{{Zeng} {et~al.}(2019){Zeng}, {Jacobsen}, {Sasselov}, {Petaev},
  {Vanderburg}, {Lopez-Morales}, {Perez-Mercader}, {Mattsson}, {Li}, {Heising},
  {Bonomo}, {Damasso}, {Berger}, {Cao}, {Levi}, \& {Wordsworth}}]{ZJS19}
{Zeng}, L., {Jacobsen}, S.~B., {Sasselov}, D.~D., {et~al.} 2019, Proc. Natl.
  Acad. Sci. USA, 116, 9723, \dodoi{10.1073/pnas.1812905116}

\end{thebibliography}

\end{document}